\documentclass[a4paper,11pt]{article}
\pdfoutput=1 

\usepackage{jcappub} 

\usepackage[T1]{fontenc} 

\newcommand{\beq}{\begin{equation}}
\newcommand{\eeq}{\end{equation}}
\newcommand{\beqarray}{\begin{eqnarray}}
\newcommand{\eeqarray}{\end{eqnarray}}
\title{ Photo-Disintegration of Heavy Nuclei at the Core of Cen A} 
\author[a]{Esha Kundu}
\affiliation[a]{Tata Institute of Fundamental Research, Homi Bhabha Road, Colaba, Mumbai 400005, India} 
\author[b]{and Nayantara Gupta}
\affiliation[b]{Raman Research Institute, Sadashiva Nagar, Bangalore 560080, India }
\emailAdd{esha.kundu@gmail.com}
\emailAdd{nayan@rri.res.in}

\abstract{
Fermi LAT has detected gamma ray emissions from the core of Cen A. 
More recently, a new component in the gamma ray spectrum from the core has been reported in the energy range of 4 GeV to tens of GeV. 
We show that the new component and the HESS detected spectrum of gamma rays from the core at higher energy have possibly a common origin in photo-disintegration of heavy nuclei. Assuming the cosmic rays are mostly Fe nuclei inside the core and their spectrum has a low energy cut-off at 52 TeV in the wind frame
 moving with a Doppler factor 0.25 with respect to the observer on earth, the cosmic ray luminosity required to explain the observed gamma ray flux above 1 GeV is found to be $1.5\times 10^{43}$ erg/sec.
}


\arxivnumber{1312.7440}

\begin{document}
\maketitle
\flushbottom
\section{Introduction}
The FRI radio-loud galaxy Cen A has been well studied throughout the electromagnetic spectrum from radio to $\gamma$ rays due to its close proximity to earth.
The gamma ray spectrum from the core of Cen A has been observed by Fermi LAT \cite{abdo1} and HESS \cite{hess}. 
The multiwavelength data provides an unique opportunity to explore the possible cosmic ray composition and the mechanism of $\gamma$ rays production at the core of Cen A.
Single zone synchrotron and synchrotron self-Compton (SSC) model have successfully explained the multiwavelenght data ranging from radio to gamma rays upto 
1 GeV \cite{abdo1}. HESS data \cite{hess} observed in the energy window of few hundred of GeV to few TeV requires hadronic interaction to justify the emission \cite{sahu,joshi}. More, recently a new component in the gamma ray spectrum from the core at GeV energy has been observed by Fermi LAT \cite{sah} which shows a hardening of gamma ray flux above 4 GeV to few tens of GeV.
The photon spectrum reveals a change of spectral index from $\Gamma \simeq $ 2.7 to 2.1 at the break energy $\sim$ 4GeV.
The observed feature is unusual as photon spectrum gets softer instead of harder at high energies. 
The new break observed in the GeV energy spectrum by Fermi LAT \cite{sah} can not be explained as the first SSC emission of the shock accelerated electrons inside the core.
However, in a recent study \cite{petro} the authors have attributed the emission above 0.4GeV to second SSC which is in the Klein Nishina regime and the TeV emission to photohadronic interactions of relativistic protons. 
The observed gamma ray emission has been explained by SSC upto 4 GeV and neutral pion decay upto TeV energy in \cite{fraija} including the possible contribution from the synchrotron radiations by the secondary muons.  

\par
In this paper we discuss about an alternative mechanism to explain the gamma ray spectrum in the GeV to TeV energy range.
It has been suggested earlier that a simple extrapolation of the new GeV component above the break can match the average HESS $\gamma$ ray spectrum \cite{sah}. We show that photo-disintegration of Fe nuclei at the core can alone explain the entire gamma ray spectrum in the GeV-TeV energy range without requiring 
multiple emission mechanisms. In this case we have a single power law spectrum to fit all the GeV-TeV gamma ray data. 

\par
 
 The composition of very high energy cosmic rays (VHECR) is not known at the core of Cen A. The cosmic rays could be mostly protons or heavy nuclei \cite{auger0}. The authors in \cite{joshi} found that pure hadronic interactions between $p$-$p$ and $Fe$-$p$ are not favored by the observational data from Pierre Auger Observatory \cite{auger1,auger2} but photo-disintegration of Fe nuclei by local radiation field can successfully explain the cosmic ray and high energy $\gamma$ ray emission from the core of Cen A. It can be an alternative to $p\gamma$ interactions at the core discussed in \cite{sahu}.    
The scenario discussed in \cite{joshi} assumes the cosmic rays are mostly Fe nuclei within the core of Cen A.

\par

In the present work we have shown that the photo-disintegration of Fe nuclei at the core can explain the new component of GeV $\gamma$ rays as well. 
A shock accelerated cosmic ray nucleus while passing through a 
radiation field may get excited by absorbing radiation. It is disintegrated to a daughter nucleus and a single proton or neutron. The daughter nucleus de-excites by emitting a photon in its rest frame, which is Lorentz boosted in the observer's frame to GeV-TeV energy.
We note that photo-disintegration can be a very efficient energy loss mechanism  at very high energy, as a result Fermi machanism may fail to accelerate the heavy nuclei. However it is possible that the acceleration
 site is not located within the intense radiation field where photo-disintegration is taking place. Within the core there could be inhomogeneity in 
magnetic and radiation fields. Shock acceleration may happen within a
 region with low photon density. The shock accelerated cosmic ray nuclei are passing through the radiation field subsequently and they are photo-disintegrated.

\section{Gamma Spectrum from the Core of Cen A}
The gamma rays  observed from the core \cite{hess,abdo1,sah}
, radio lobes \cite{abdo2} and kiloparsec scale jets of Cen A provide interesting probes for testing different emission mechanisms \cite{gupta,kachel,hard} at these emission sites.
\par
 It was suggested earlier that the photo-disintegration of heavy nuclei at the core of Cen A could be the origin \cite{joshi} of the gamma ray flux  observed by HESS \cite{hess}. In this work we add the new gamma ray data from Fermi LAT \cite{sah} to the spectral energy distribution given in \cite{abdo1}. With the combined data from Fermi LAT \cite{abdo1,sah} and HESS \cite{hess} we show that the photo-disintegration of heavy nuclei at the core remains a viable mechanism of gamma ray production.
\par 
 The wind medium at the core is moving relativistically with Lorentz factor $\Gamma_j$. As discussed in \cite{joshi} the gamma rays are detected only if the beaming of the gamma ray emission in the observer's frame is along the observer's line of sight. 
Depending on the angle of observation of the gamma ray emission $\theta_{ob}$, the Doppler factor of the wind medium is $\delta_D=\Gamma_j^{-1}(1-\beta_j
 \, cos{\theta_{ob}})^{-1}$, where $\beta_j$ is the dimensionless speed of the wind rest frame with respect to the observer on earth. 
\par

We revisit the photo-disintegration model giving the equations relevant for our study. The rate of photo-disintegration process given in \cite{stecker,anch3} is
\beq
R_{photdis,A}=\frac{c\pi \sigma_0 \epsilon_0'\Delta}{4\gamma_p^2} \int_{\epsilon_0'/2\gamma_p} ^\infty  \frac{dn(x)}{dx}\frac{dx}{x^2}.
\label{rate_photodis}
\eeq
The value of the cross-section normalization constant is $\sigma_0=1.45A$ mb, the central value of GDR is $\epsilon_0'=42.65 A^{-0.21}(0.925A^{2.433})$ MeV for $A>4 (A \leq 4)$ and width of the GDR is $\Delta=8$ MeV. 
$\gamma_p=E_{A}/(A m_p)$, represents the Lorentz factor of each nucleon.
The spectral energy distribution (SED) observed on earth, denoted by ${\epsilon_{\gamma}^o}^2\frac{dN_{\gamma}^o(\epsilon_{\gamma}^o)}{d\epsilon_{\gamma}^odt^odA}$(MeV cm$^{-2}$ sec$^{-1})$, is shown with black solid curve in figure \ref{fig:2}. 
The photon density per unit energy at the core $\frac{dn(x)}{dx}$ is related to the observed SED as follows 
\beq
4\pi R_w^2  c \frac{dn(x)}{dx}=4\pi D^2\delta_D^{-p}\frac{dN_{\gamma}^o(\epsilon_{\gamma}^o) }{d\epsilon_{\gamma}^o dt^o dA}
\label{phot_den}
\eeq
where, $p=n+\alpha+2$ and $n=2$ for the observed beamed emission \cite{gis}. 
The observed energy of the low energy photons is related to their energy 
in the wind rest frame by the Doppler factor $\delta_D$ of the wind medium, $\epsilon_{\gamma}^o=\delta_D x$. The blob size $R_w=R\delta_D$ measured in wind rest frame becomes $R$ in the observer's frame.  
\par 
In \cite{abdo1} various models have been fitted to the SED with different values of bulk Lorentz factor $\Gamma_j$ and $\delta_D$. 
They showed Fermi and HESS data can not be fitted together with a single zone SSC model. We note that at that time the new Fermi data \cite{sah} was not present.
\par
In eq.(\ref{phot_den}) it is to be noted that a small change in the value of $\delta_D$ leads to large change in photon density at the core as it is proportional to $\delta_D^{-4-\alpha}$. 
 The photon energy flux ${\epsilon_{\gamma}^o}^2\frac{dN_{\gamma}^o(\epsilon_{\gamma}^o)}{d\epsilon_{\gamma}^odt^odA} \propto {\epsilon_{\gamma}^o}^{-\alpha}$, where the spectral index $\alpha$ takes different values in different energy regimes as shown in figure \ref{fig:2}.
\par

\begin{figure}[tbp]
\centering
\includegraphics[width=\textwidth,origin=c]{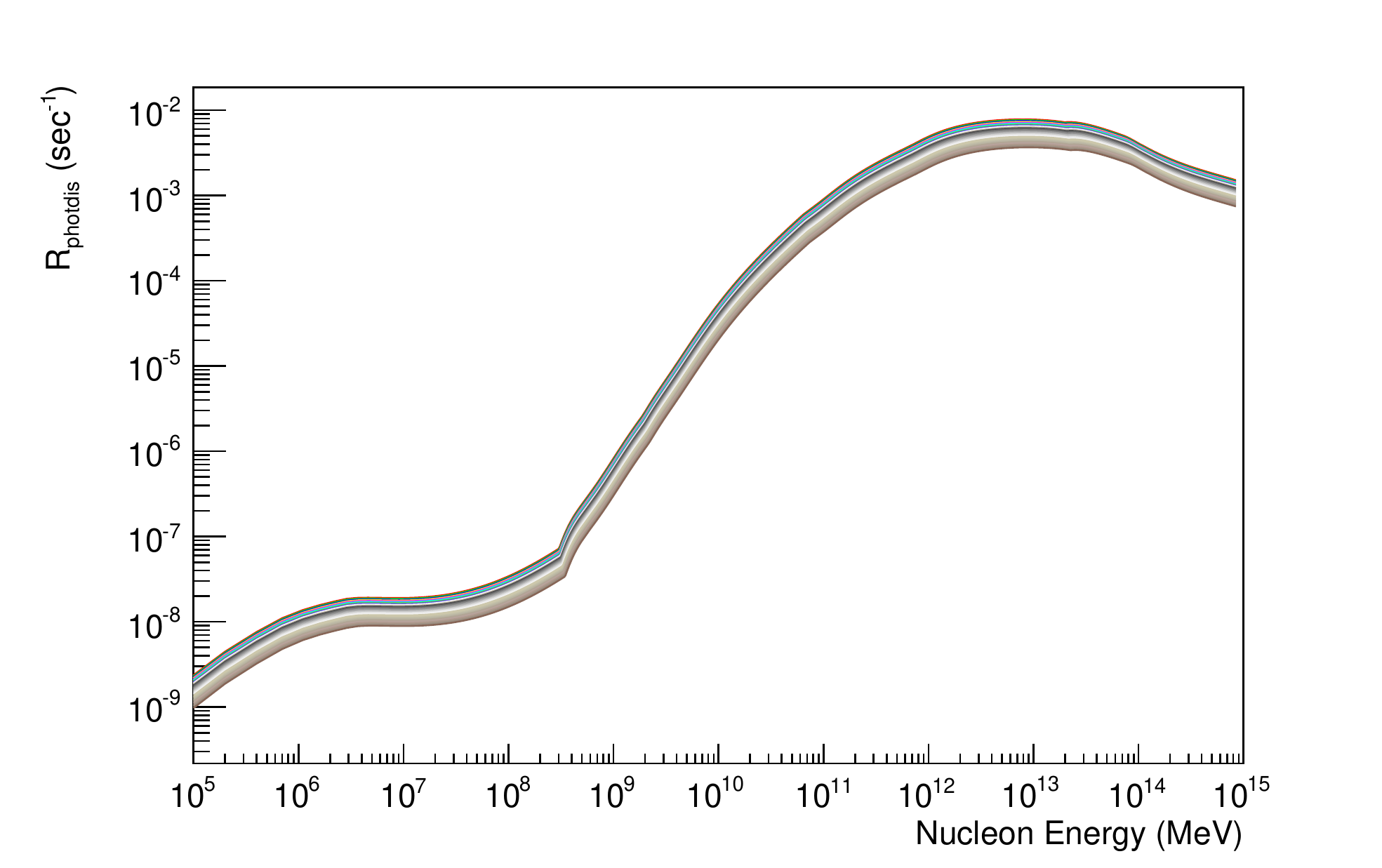}
\caption{\label{fig:1}
Photo-disintegration rate $R_{photdis}$ of heavy nuclei as a function of nucleon energy for mass number A = 56 to A = 30 in descending order in the wind rest frame for the values of the parameters given in \cite{abdo1} also shown in third column of our table \ref{tab1}.}
\end{figure}

\begin{figure}[tbp]
\centering
\includegraphics[width=\textwidth,origin=c]{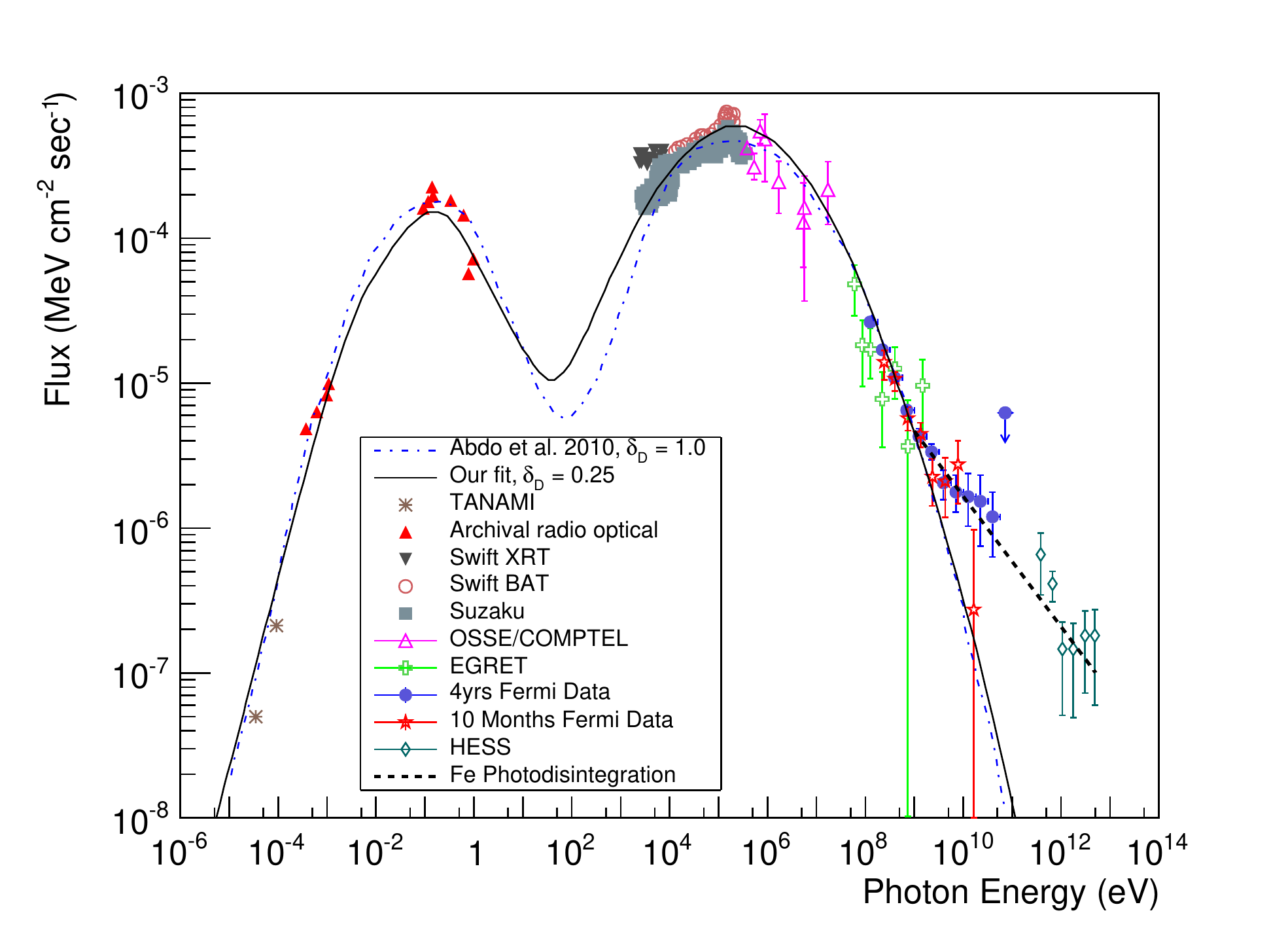}
\caption{\label{fig:2}
Photon flux in MeV cm$^{-2}$ sec$^{-1}$, data points from different experiments compiled in \cite{abdo1}, 4 years Fermi data shown from \cite{sah}. 
SED for $\delta_D=0.25$ fitted using tool from \cite{tram1,tram2,mass}.}
\end{figure}

With the size of the emission region of Cen A $R=3\times 10^{15}$ cm, Doppler factor $\delta_D=1$ \cite{abdo1} and the distance of Cen A to us $D=3.4$ Mpc the photo-disintegration rates in the wind rest frame for nuclei with mass number A = 56 to A = 30 are shown in figure \ref{fig:1} in descending order.

The rate of photo-disintegration is almost constant for all nuclei with mass number A = 56 to A = 30 in the energy window of 1-100 TeV in the wind rest frame. In the energy interval of 1-100 TeV the rate for A = 56 is $2\times10^{-8}$ sec$^{-1}$ which decreases to $9\times10^{-9}$ sec$^{-1}$ for A = 30. 
As the rate decreases with decreasing values of A, the major contribution to the observed $\gamma$ ray flux comes from photo-disintegration of nuclei with higher values of A. 
\par
In the process of photo-disintegration of heavy nuclei protons and neutrons are 
 stripped from the parent nuclei with equal probability.
 This model has been discussed in detail in earlier papers \cite{stecker,anch3}.  The gamma ray flux expected is related to the number of parent nuclei per unit nucleon energy per unit time measured in the wind rest frame
$\frac{dN_{A}}{dE_N dt}(E_N)=C_A E_N^{-\Gamma}$ as follows:
\beq
\frac{d\phi_{\gamma}(E_{\gamma})}{dE_{\gamma} dt}=\frac{R_w}{\beta c}
\frac{\bar n_{A} m_N}{2 \bar E'_{\gamma,A}} C_A \Big(\frac{m_N  E_{\gamma}}{2 \bar E'_{\gamma,A}}\Big)^{-\Gamma} R_{photdis,A} 
\label{phot_dis}
\eeq
where $\beta=v/c $ dimensionless speed of the high energy nuclei. 
Eq.(\ref{phot_dis}) can be written when the following conditions are satisfied.
The mean free path length of photo-disintegration is longer than the size of the blob. This is true in our case. The GeV-TeV photons considered in our work are produced from nucleons of energy below 1.2 PeV. From the rates shown in figure \ref{fig:1} for $\delta_D=1$ one can see that the mean free path lengths of photo-disintegration are much longer than the size of the blob $3\times 10^{15}$ cm below 1.2 PeV. The second condition is the $\gamma$ ray spectrum is assumed to be monochromatic in the rest frame of a nucleus. The average energy of the gamma ray photon is denoted by $\bar E'_{\gamma,A}$. For a nucleus with mass number A, $\bar n_{A}$ represents the mean $\gamma$ ray multiplicity and $m_{N}$ is the rest mass of each nucleon.
For Fe nuclei we have used $\bar E'_{\gamma,56} =$ 2-4 MeV and $\bar n_{56} = $ 1-3. 
The photon flux at energy $E_{\gamma}$ is produced from cosmic rays of per nucleon energy $E_N=\frac{m_N  E_{\gamma}}{2 \bar E'_{\gamma,A}}$.
The quantity $\frac{\bar n_{A} m_N}{2 \bar E'_{\gamma,A}}$ on the right hand side of eqn.(\ref{phot_dis}) is dimensionless, hence both sides of this equation have the same dimension of per unit energy per unit time. 
\par
The new Fermi data \cite{sah} starts from 4 GeV and the HESS data has the maximum energy 5 TeV. 
We have fitted the gamma ray data starting from 1 GeV to 5 TeV with a single power law spectrum.
The gamma rays of energy 1 GeV to 5 TeV in the observer's frame are produced from Fe cosmic ray nuclei with energy per nucleon $E_N$ ranging from 234 GeV to 1.17 PeV assuming $\bar E'_{\gamma,56} =$ 2 MeV and Doppler factor $\delta_D=1$. 
Within this energy range the rate of photo-disintegration increases from $5\times 10^{-9}$ to $1.2\times 10^{-6}$ sec$^{-1}$, but it is almost constant over the cosmic ray nucleon energy of 1$-$100TeV. The average rate over this energy range is $R_{photdis}=2\times10^{-8}$ sec$^{-1}$ in the wind rest frame.

The gamma ray spectrum above 250 GeV \cite{hess} in the observer's frame is
\beq
\frac{d\phi_{\gamma}(E^o_{\gamma})}{dE^o_{\gamma} dt^o dA}=\frac{1}{4\pi D^2 \delta_D^2}\frac{d\phi_{\gamma}(E_{\gamma})}{dE_{\gamma} dt}=2.45\times 10^{-13} \Big(\frac{E_{\gamma}^o}{1 TeV}\Big)^{-2.73} TeV^{-1} cm^{-2} sec^{-1}.
\label{phot_dis_obs}
\eeq
We have assumed the same spectrum of gamma rays for the entire GeV-TeV energy range. Very high energy $\gamma$ rays observed by HESS has a spectral index of $2.73\pm0.45_{stat}\pm0.2_{sys}$. We have used $\gamma$ ray spectral index of 2.45 \cite{joshi} which is within the range of error in the spectral index of HESS detected $\gamma$ ray. 
The normalization constant of the cosmic ray spectrum ($C_A$) is calculated using eq.(\ref{phot_dis}) and eqn.(\ref{phot_dis_obs}) with  $\Gamma=2.45$ and photon multiplicity $\bar n_{56}=$3. 
The high energy $\gamma$ ray flux obtained from photo-disintegration of Fe nuclei is shown in figure \ref{fig:2} with black dashed line. 
The new component in the gamma ray data observed by Fermi LAT \cite{sah} and the HESS data \cite{hess} are well fitted with the gamma ray flux obtained from photo-disintegration of Fe nuclei. 

The required cosmic ray luminosity is found to be very high ($5.9\times 10^{46}$ erg/sec) for the production of $\gamma$ rays of energy more than 1 GeV using the values of the parameters given by Abdo et al.\cite{abdo1}, also tabulated in third column of our table \ref{tab1}.
\par
 For Cen A the estimated jet angle $\theta_{ob}$ lies in the range of $15-80^{o}$\cite{rous,tin} and Doppler factor $\delta_D$ in between 0.12 - 3.7 \cite{abdo1,rous,kotha,chi}.
We have fitted the photon data below 1 GeV with the SED fitting tool from \cite{tram1,tram2,mass} for a lower Doppler factor $\delta_D = 0.25$ considering synchrotron and SSC emissions, shown with solid line in figure \ref{fig:2}. We note that the sizes of the emission region for the $\delta_D=1$ and $\delta_D=0.25$ cases are $3\times 10^{15}$ cm and $5.8\times 10^{15}$ cm respectively.
The energy of the photons responsible for the photo-disintegration of Fe nuclei are calculated using the expression $x=\epsilon'_0/2 \gamma_p$ where $\epsilon_0'=42.65 \times 56^{-0.21}$ MeV.
 With the values of the parameters obtained from our fit (shown in fourth column of table \ref{tab1}) the photo-disintegration rate increases (see our eqn.(\ref{rate_photodis}) and eqn.(\ref{phot_den})) by a factor of $(1/0.25)^5\times (3/5.8)^2=274$ where $\alpha=-1$ for the keV photons and the size of the blob in the wind rest frame $R_w=R \delta_D$.
In eqn.(\ref{phot_dis}) we have multiplied with the size of the blob $R_w$ on the right hand side and in eqn.(\ref{phot_dis_obs}) the energy and time are corrected with $\delta_D$ to get the gamma ray flux in the observer's frame. 
Finally the normalization constant reduces by a factor of $(1/0.25)^6\times (3/5.8)=2119$.
 Moreover, for $\delta_D = 0.25$, the $\gamma$ rays of energy 1 GeV in the observer's frame are produced from cosmic ray Fe nuclei with energy per nucleon $E_N = 4 \times 234.5$ GeV = 938 GeV in the wind rest frame.
After including all the effects of lowering the Doppler factor from $\delta_D=1$ to $\delta_D=0.25$, the required cosmic ray power reduces to $1.5 \times 10^{43}$ erg/sec, which is comparable to the jet luminosity $\sim 10^{43}$ erg/sec. 
Thus photo-disintegration of heavy nuclei remains as a possible mechanism for the production of the observed GeV-TeV $\gamma$ rays inside the core of Cen A. 
For photo-disintegration to be a dominant process at the core the primary cosmic rays have to be mostly heavy nuclei. 
\par
 Our calculation of the luminosity of the cosmic rays for the production of gamma rays above 1 GeV assumes a low energy cut-off in the cosmic ray spectrum near $0.938\times 56$ TeV = 52 TeV in the wind rest frame for $\delta_D = 0.25$. 
In the case of $\delta_D = 1$ this low energy cut-off occurs at $0.234\times 56$ TeV = 13 TeV. 
We have introduced this cut-off in the spectrum of Fe cosmic ray nuclei, which has no physical interpretation at present, to reduce the required cosmic ray luminosity.

\begin{center}
\begin{tabular}{| c | c | c | c |}
 \hline
 Parameter  & Symbol & Abdo et al. 2010 Model & Our Fit  \\ 
 \hline
 Comoving blob Size (cm) & R  & 3 $\times$ $10^{15}$ & $5.8 \times 10^{15}$  \\ 
 Bulk Lorentz factor & $\Gamma_j$ & 7.0 & 7.0 \\ 
 Doppler factor & $\delta_D$  & 1.0 & 0.25  \\ 
 Jet angle & $\theta_{ob}$ & $30^{o}$ & $65^{o}$ \\
 Magnetic field (G)& B & 6.2 & 33 \\
 Variability time scale (sec)  & $t_{var}$ & 1.0 $\times$ $10^5$ & $1.0 \times 10^5$  \\
 \hline
 Low energy electron spectral index &  $p_1$  &  1.8 & 1.8 \\
 High energy electron spectrla index &  $p_2$ & 4.3 & 4.3 \\
 Minimum electron Lorentz factor & $\gamma_{min}$ &  3 $\times$  $10^2$ & $1.3 \times  10^2$ \\ 
 Maximum electron Lorentz factor & $\gamma_{max}$ & 1 $\times$  $10^8$  & $1.0 \times 10^8$  \\
 Break electron Lorentz factor & $\gamma_{brk}$  &  8 $\times$  $10^2$ & $1.04 \times  10^3$  \\
 \hline
 \end{tabular}
 \begin{table}[htbp]
 \caption{ \label{tab1}
 Parameter values of Abdo et al. 2010 model\cite{abdo1} with $\delta_D = $1. and our fit for $\delta_D = $0.25.}
\end{table}
\end{center}

\section{High Energy Cosmic Rays and Antineutrinos from the Core}

\begin{figure}[tbp]
 \centering
 \includegraphics[width=0.65\textwidth,origin=c]{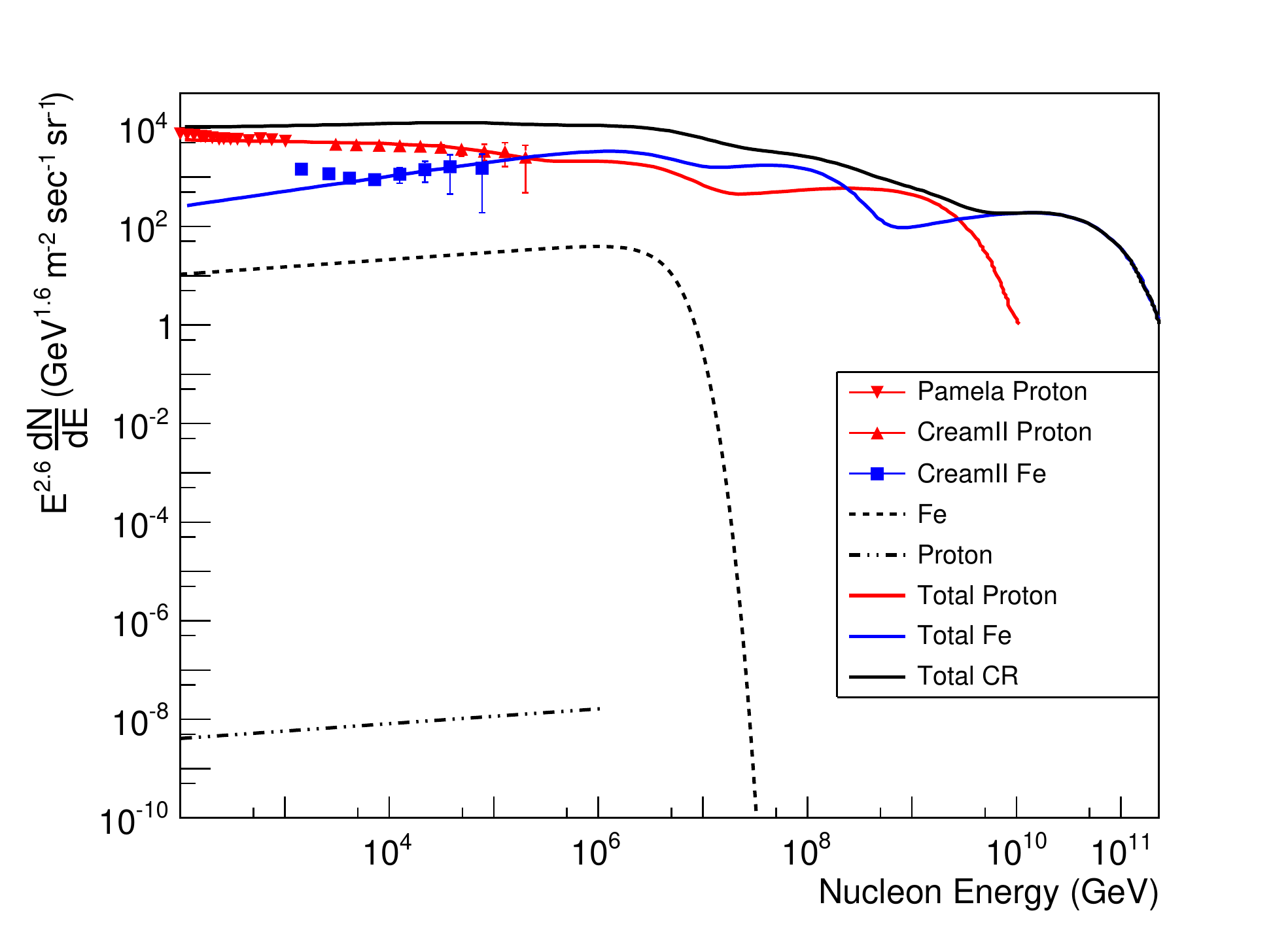}
 \caption{\label{fig:3}
Cosmic ray Fe nuclei and proton fluxes injected from Cen A, calculated using model parameters given in \cite{abdo1} and in third column of our table \ref{tab1}, are compared with cosmic ray data points from PAMELA \cite{pamela} and CREAMII \cite{cream}. Total cosmic ray flux, Fe and proton fluxes from figure 4 (left side) and Global fit given in Table III of \cite{gaisser_new}.}
\end{figure}

We calculate the cosmic ray Fe nuclei and proton fluxes escaping from the core of Cen A using the model parameters given in \cite{abdo1}, also shown in third column of our table \ref{tab1}. 
In photo-disintegration of heavy nuclei, neutrons and protons are stripped
from the parent nuclei with almost equal probability.
 The average number of gamma rays emitted from the de-exciting daughter nuclei
 or the photon multiplicity $\bar n_{56}$ and the probability of a proton emission $1/2$ allows us to relate the gamma ray flux with the proton or neutron flux in the daughter rest frame. After Lorentz transformation the fluxes are related \cite{anch3} in the observer's frame in this way

\beq
\frac{dF_{p(n)}}{dE_{p(n)}} (E_{p(n)}) = \frac{\bar E'_{\gamma,56}}{2 m_{p(n)} \bar n_{56}}  \frac{dF_{\gamma}}{dE_{\gamma}} (E_{\gamma} = \frac {E_{p(n)} \bar E'_{\gamma,56}}{m_{p(n)}}) 
\label{p_flux}
\eeq
Where $\frac{dF_{p(n)}}{dE_{p(n)}}$ and $\frac{dF_{\gamma}}{dE_{\gamma}}$ represent the proton(neutron) and gamma ray 
flux in the observer frame and $m_{p(n)}$ is the rest mass of proton(neutron). 
The probability of $p\gamma$ interaction is low for the protons below 1 PeV.
We find the neutrino flux from photo-hadronic interactions of these protons is much lower than the neutron decay antineutrino flux.
The cosmic ray fluxes of Fe and protons escaping from the core of Cen A are shown in figure \ref{fig:3} and compared with observed data on diffuse cosmic rays from PAMELA \cite{pamela} and CREAM II \cite{cream}. 

\begin{figure}[tbp]
 \centering
  \includegraphics[width=0.65\textwidth,origin=c]
{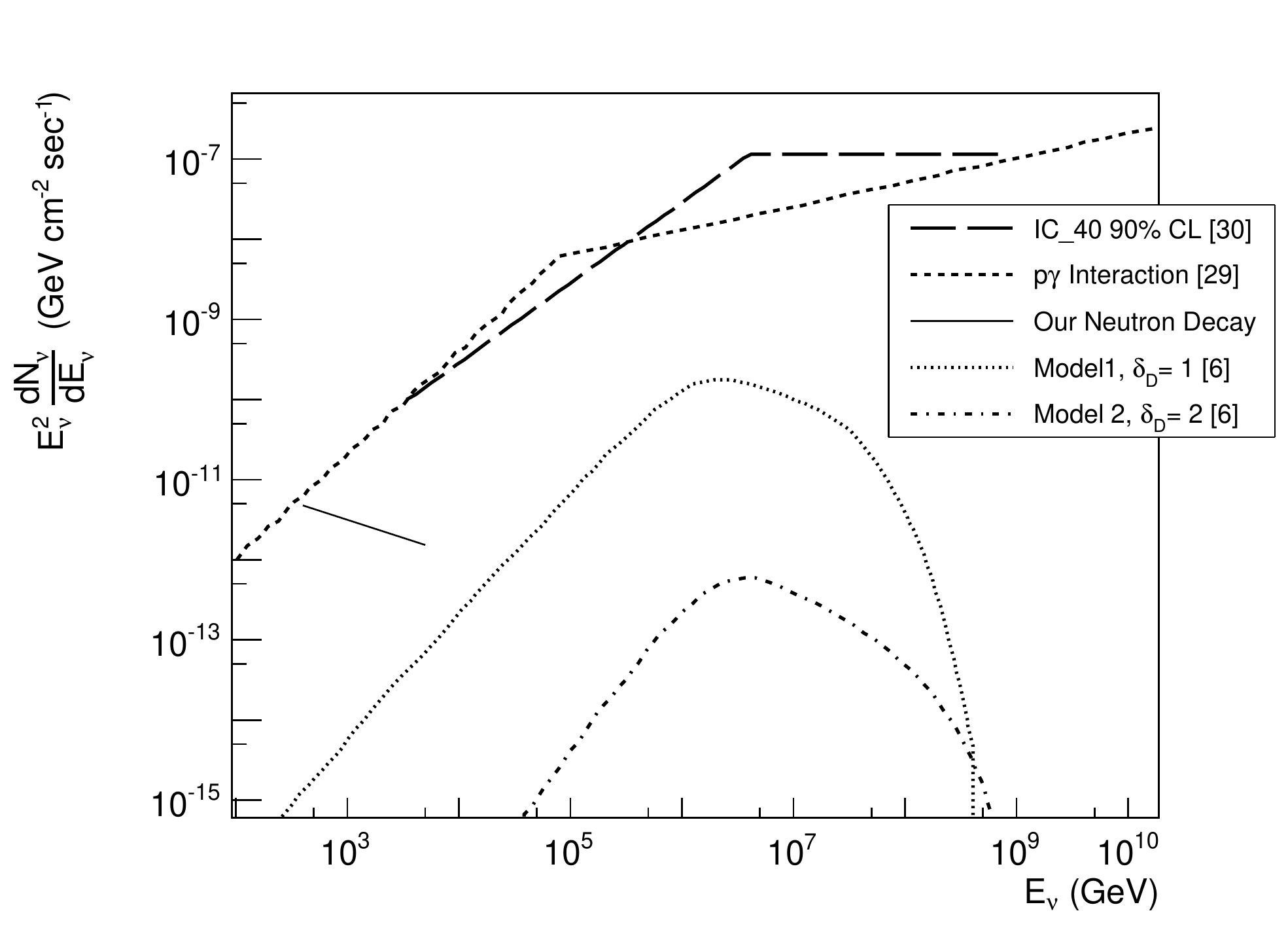}
 \caption{\label{fig:4}
Our estimated antineutrino flux from neutron decay (solid line) calculated using our eqn.(\ref{neutrino_n_decay}) has been compared with other model predictions. Neutrino flux from $p\gamma$ interaction (small dashed line) from \cite{cuoco}, Model 1 and Model 2 from \cite{petro} and IceCube upper limit with $90\%$ C. L. from \cite{abbasi} are shown.}
\end{figure}

The stripped neutrons can be emitted in any directions as VHE Fe nuclei are boosted in all 4$\pi$ sr in the core. 
Neutrons which are moving toward us with energy $\sim$ 10$^{6}$ GeV can decay to antineutrinos. 
If the spectral index of photon flux produced by photo-disintegration of Fe nuclei is $\alpha$, and the 
average energy of antineutrino is $\epsilon_0 = 0.48$ MeV, then in the observer's frame the antineutrino flux $\frac{dF_{\bar\nu}}{dE_{\bar\nu}}$ 
is related to the gamma ray flux $\frac{dF_{\gamma}}{dE_{\gamma}}$ as follows:
\beq
\frac{dF_{\bar\nu}}{dE_{\bar\nu}}(E_{\bar \nu}) = \frac{1}{\alpha} \Bigg({\frac{2\epsilon_0}{\bar E'_{\gamma,56}}}\Bigg)^{\alpha -1}   \frac{1}{2 \bar n_{56}}   \frac{dF_{\gamma}}{dE_{\gamma}} (E_{\gamma} = E_{\bar \nu})  
\label{neutrino_n_decay}
\eeq
The antineutrino flux from neutron decay is calculated using eq. (\ref{neutrino_n_decay}) and plotted with solid line in figure \ref{fig:4}. 
 We note that in eq. (\ref{neutrino_n_decay}) the antineutrino energy is equal to the gamma ray energy, $E_{\gamma} = E_{\bar\nu}$ . 
We have shown the neutron decay antineutrino flux in the energy window for which HESS $\gamma$ ray data is available.

\section{Discussion and Conclusions}

The latest $\gamma$ ray data from the core of Cen A \cite{sah} shows a new component above 4 GeV which can not be explained with the single zone first SSC spectrum given in \cite{abdo1}.  We show that this new Fermi LAT data and the HESS data may have a common origin in photo-disintegration of Fe nuclei. Our calculated spectrum is shown with black dashed line in figure \ref{fig:2}.
The cosmic ray power required to explain the gamma ray emission above 1 GeV is $1.5\times 10^{43}$ erg/sec.
This is calculated assuming a cut-off in the Fe cosmic ray spectrum at 52 TeV in the wind rest frame for $\delta_D = 0.25$. Our model requires this 
ad-hoc low energy cut-off in the cosmic ray Fe spectrum to reduce the required luminosity in cosmic rays.
\par
 The cosmic rays emitted from Cen A are deflected in the extragalactic magnetic field. They are assumed to be isotropically distributed around their source after many deflections. 
In that case the cosmic ray energy flux at a distance of $D=3.4$ Mpc from the source is $L_{esc,CR}/4 \,\pi \, D^2$ erg/cm$^2$/sec{\bf /sr}, where $L_{esc,CR}$ erg/sec is the luminosity of the cosmic rays escaping from the core of Cen A.
After entering our Galaxy the cosmic rays are further deflected in the Galactic magnetic field and we expect them to become isotropic around the earth after many deflections.
The propagated flux of cosmic rays is lower than the injected flux due to diffusion and interaction losses inside our Galaxy.

The cosmic ray Fe and proton fluxes injected from Cen A core are shown with dashed and dash double dotted lines in figure \ref{fig:3}. 
 We note that the effect of the strong Galactic magnetic field on the cosmic rays has to be studied in detail for a more precise estimation of the cosmic ray flux received from Cen A.
 The observed total cosmic ray, Fe and proton fluxes from figure 4 (left side) of \cite{gaisser_new} are shown with solid lines to compare with our calculated fluxes.
As the photo-disintegration rate increases rapidly with energy, at very high energy the escaping Fe flux falls sharply. The proton flux produced in stripping of parent heavy nuclei is very low. The maximum energy of the cosmic rays are constrained from the acceleration mechanism and the rate of energy loss.    
\par
The antineutrino flux produced from the decay of the stripped neutrons is shown with solid line in figure \ref{fig:4}. It is found to be much less than the neutrino flux estimated in \cite{cuoco} considering $p\gamma$ interaction followed by the decay of charged pion, shown with small dashed line in figure \ref{fig:4}.  Their neutrino flux has a break around 100 TeV beyond which the flux flattens to a spectral index of $-1.7$.
They estimated above 100TeV neutrino event rate in IceCube to be 0.35 yr$^{-1}$ to 0.56 yr$^{-1}$ which can result in a few detectable events from Cen A in 5 years. 
\par
$pp$ interaction has been tested by the authors in \cite{halzen} as a possible source of neutrino production from Cen A and M87. They have shown that 
for a FRI density of $n \simeq 8 \times 10^{4}$ Gpc$^{-3}$ within a horizon of $R \sim 3$Gpc, where the galaxies are assumed to be distributed 
uniformly, the total neutrino flux from all sky is one tenth of Waxman-Bahcall flux for a neutrino spectrum having spectal index of $-2$.
Varying the power-law indices between $-2$ and $-3$ they have estimated diffuse neutrino event rate per km$^{2}$ year from the northern sky between 19 and 0.5.
In our scenario antineutrino events are not expected to be detected in IceCube as their flux is sharply falling off at TeV energy.
\par
We have also shown the neutrino fluxes predicted in \cite{petro} from photo-hadronic interactions ($p \, \gamma$) in figure \ref{fig:4}. 
They have considered two sets of model parameters denoted as Model 1 and Model 2, displayed in their Table 3. Their Model 1 and Model 2 use $\delta_D=1$ and $\delta_D=2$ respectively.
In both models they have used a high value for the minimum proton Lorentz factor to reduce the required luminosity in cosmic ray protons. In their models a peak in the neutrino flux is expected near PeV energy. Thus the different scenarios of neutrino production are easily distinguishable.
\par
IceCube upper limit with $90\%$ confidence limit\cite{abbasi} for neutrino detection is shown in figure \ref{fig:4} with long dashed line.
\par
If photo-disintegration is the underlying mechanism of gamma ray production at the core of Cen A then the core contains mostly heavy nuclei and we do not expect to detect any neutrino or antineutrino event from the core in IceCube. 

\end{document}